\documentstyle[aps,multicol]{revtex}

\draft

\begin{document}

\title{Comment on ``Validity of Feynman's prescription of disregarding
the Pauli principle in intermediate states''}
\author{R M Cavalcanti\footnote{E-mail: rmoritz@fma.if.usp.br}}
\address{Instituto de F\'{\i}sica, Universidade de S\~ao Paulo,
Cx.\ Postal 66318, 05315-970 S\~ao Paulo, SP, Brazil}
\maketitle

\begin{abstract}

In a recent paper Coutinho, Nogami and Tomio [Phys.\ Rev.\ A
{\bf 59}, 2624 (1999)] presented an example in which, they
claim, Feynman's prescription of disregarding the Pauli
principle in intermediate states of perturbation
theory fails. We show that, contrary to their claim,
Feynman's prescription is consistent with the exact solution
of their example. 

\end{abstract}



\begin{multicols}{2}

Feynman's prescription of disregarding the exclusion
principle in intermediate states of perturbation theory
is based in his observation \cite{Feynman}
that all virtual processes that violate
that principle (formally) cancel out order by order in
perturbation theory. 
However, Coutinho, Nogami and Tomio have recently
presented an example \cite{Coutinho} in which Feynman's
prescription seems to fail. They calculated (in second order
perturbation theory) the energy shift $W\equiv E(\lambda)-E(0)$
of the ground state of the
one-particle sector of the one-dimensional bag-model
caused by the potential $V(x)=\lambda x$.
The calculation was performed using both prescriptions, i.e.,
either excluding virtual transitions to occupied states,
in accordance with the Pauli principle (method I), or including
such transitions, as suggested by Feynman (method II). 
It turns out that the result depends on the prescription used;
in the massless case, for instance, 
$W_{\rm I}\ne 0$ and $W_{\rm II}=0$.
In view of this discrepancy, Coutinho et al.\ suggested that one
should abandon Feynman's prescription and remain
faithful to the Pauli principle in every step of
the calculation. 

The purpose of this Comment is to
show that the opposite alternative is the correct one.
More precisely, we show that $W_{\rm II}$ agrees
(while $W_{\rm I}$ does not) with the exact value of $W$ 
in the massless case. 

To calculate $W$ we need to solve the Dirac equation 
\begin{equation}
\label{Deq1}
(H_0+\lambda x)\,\psi(x)=\epsilon\,\psi(x),
\end{equation} 
where $H_0$ is the Hamiltonian of the bag model
(see \cite{Coutinho} for its definition). With the 
conventions of ref.\ \cite{Coutinho}, 
Eq.\ (\ref{Deq1}) can be written explicitly 
in the massless case as
\begin{equation}
\label{Deq2}
\left(\begin{array}{cc}
\lambda x-\epsilon & -\partial_x \\
\partial_x & \lambda x-\epsilon
\end{array}\right)
\left(\begin{array}{c}
u \\ v
\end{array}\right)
=\left(\begin{array}{c}
0 \\ 0
\end{array}\right)\qquad(|x|\le a),
\end{equation}       
with $u$ and $v$ subject to the boundary conditions
$u(\pm a)=\mp v(\pm a)$.
Defining $w_{\pm}=u\pm iv$, we can
rewrite (\ref{Deq2}) as
\begin{equation}
\partial_xw_{\pm} \mp i(\lambda x-\epsilon)\,w_{\pm} = 0, 
\end{equation}
whose general solution is
\begin{equation}
w_{\pm}(x)=C_{\pm}\,e^{\pm i\left(\frac{\lambda}{2}x^2
-\epsilon x\right)}.
\end{equation}
The boundary conditions on $u$ and $v$ turn into
$w_{+}(\pm a)=\mp iw_{-}(\pm a)$. This gives two
relations between $C_{+}$ and $C_{-}$:
\begin{eqnarray}
C_{+}&=&iC_{-}\,e^{-i\left(\lambda a^2
+2\epsilon a\right)}, 
\label{r1}
\\ 
C_{+}&=&-iC_{-}\,e^{-i\left(\lambda a^2
-2\epsilon a\right)}. 
\label{r2}
\end{eqnarray}
Dividing (\ref{r1}) by (\ref{r2}) and solving for
$\epsilon$ gives
\begin{equation}
\epsilon=(2n+1)\,\frac{\pi}{4a}\qquad(n=0,\pm 1,\pm 2,\ldots).
\end{equation}
Since the energy levels do not depend on $\lambda$,
the energy shift $W$ is zero. 
This is precisely the result obtained
by Coutinho et al.\ using Feynman's prescription
in perturbation theory.

\vspace{3mm}

The author acknowledges the financial support from FAPESP.


\end{multicols}

\end{document}